\documentclass[pra,twocolumn,showpacs,amsmath,amssymb]{revtex4}

\newcommand{\subsize}
{\scriptsize}

\newcommand{\subsubsize}
{\tiny}

\newcommand{\sub}[1]
{_\textrm{\subsize #1}}

\newcommand{\subsub}[1]
{_\textrm{\subsubsize #1}}

\newcommand{\gammafktB}[2]
{\frac{\Gamma\big(\frac{#1}{2}\big)}{\Gamma\big(\frac{#2}{2}\big)}}

\newcommand{\BE}[1]
{Bose-Einstein}

\begin{document}

\title{Correlated $N$-boson systems for arbitrary scattering length}

\author{O.~S\o rensen}
\author{D.~V.~Fedorov}
\author{A.~S.~Jensen}

\affiliation{Department of Physics and Astronomy, University of
  Aarhus, DK-8000 Aarhus C, Denmark}

\date{\today}

\begin{abstract} 
  We investigate systems of identical bosons with the focus on
  two-body correlations and attractive finite-range potentials.  We
  use a hyperspherical adiabatic method and apply a Faddeev type of
  decomposition of the wave function.  We discuss the structure of a
  condensate as function of particle number and scattering length. We
  establish universal scaling relations for the critical effective
  radial potentials for distances where the average distance between
  particle pairs is larger than the interaction range.  The
  correlations in the wave function restore the large distance
  mean-field behaviour with the correct two-body interaction.  We
  discuss various processes limiting the stability of condensates.
  With correlations we confirm that macroscopic tunneling dominates
  when the trap length is about half of the particle number times the
  scattering length.
\end{abstract}


\maketitle

\section{Introduction}

\label{sec:introduction}

Condensation of a macroscopic number of bosons in the same quantum
state was predicted many years ago \cite{bos24}.  Much later this was
experimentally achieved in the laboratory for dilute systems of alkali
gases \cite{and95,bra95,dav95}. The average properties of these gases
are accounted for by the Gross-Pitaevskii equation \cite{edw95,bay96}.
Exhaustive reviews of the theoretical developments after the
experimental breakthrough can be found in \cite{dal99,pet01}.

Degrees of freedom beyond the mean-field are crucial for the stability
of the condensates, e.g., recombination into bound dimer and trimer
cluster states \cite{nie99,esr99,bed00}. The importance of such
correlations is revealed in recent experiments \cite{rob01,don01}.
One example is the collapse of a Bose gas with large scattering length
\cite{don01} where the lack of atoms in the condensate challenged the
mean-field description in terms of the time-dependent Gross-Pitaevskii
equation \cite{adh02b,adh02d}.

The mean-field description is valid for $n|a_s|^3\ll1$, where $n$
is the density and $a_s$ is the two-body $s$-wave scattering length,
when the particles on average are outside the interaction volume of
the order of scattering length, $a_s$, to the third power
\cite{pet01}.  The mean-field method neglects all correlations and
thus breaks down at larger densities where correlations become
important.  Going beyond the mean-field is often very complicated as
exemplified by Jastrow theory \cite{jas55,cow01,mou01}, which leads to
high-dimensional equations.  A later formulation is contained in the
Faddeev-Yakubovski{\u\i} equations \cite{fad61,yak67} where the wave
function is expressed in terms of components describing the asymptotic
behaviour of all kinds of clusters.

Comparison of different models is not always straightforward, since
different degrees of freedom are treated and the two-body interactions
must be renormalized accordingly. For example, using realistic
potentials in selfconsistent mean-field calculations leads to
disastrous results because the Hilbert space does not include
correlations \cite{esr99b} as needed to describe both the short- and
long-range asymptotic behaviour. In Gross-Pitaevskii calculations the
$\delta$-function interaction is renormalized to give the correct
scattering length in the Born approximation \cite{pet01}. However,
this substitution is only valid in the low-density limit.  When
correlations are included a different, and more realistic, interaction
must also be used.  Furthermore, only average properties can be
described in the mean-field approximation.  Thus comparisons of
correlation dependent quantities are meaningless.

Five years ago an interesting alternative study of a condensate was
formulated in terms of hyperspherical coordinates without any two-body
correlations \cite{boh98}.  Using the same coordinate system a
theoretical frame for describing correlations was given soon after
\cite{bar99a}.  Detailed three-body calculations with zero total
angular momentum were recently performed in the same framework
\cite{blu02c}. Here the scattering length is varied and excited
three-body states and any number of bound two-body molecular states
are allowed.  The claim in \cite{blu02c} is that higher-lying
Bose-Einstein condensed states in a trap of length $b\sub t$ do not
collapse when $N|a_s|/b\sub t\gtrsim0.5$ as otherwise indicated by
experiments \cite{don01}.  The recombination takes place at distances
several times the scattering length.  They conjecture that the
properties for $N>3$ are quantitatively similar to these three-body
results.  Another study in the same framework investigates the
model-dependence of the three-body energy and finds that only the
large scattering length enter \cite{jon02}.  They also indicate that
the energy is insensitive to possible higher-order correlations for
systems with many particles in a trap.

In a further development using the adiabatic hyperspherical expansion
we formulated a method to describe two-body correlations in many-boson
systems \cite{sor02,sor02b}. This method is a novel attempt to
describe correlated systems of low density. The formulation heavily
relies on an additive set of components of the wave function as in the
Faddeev decomposition but in contrast to the Jastrow multiplicative
formulation.  The numerical studies were limited to Bose-Einstein
condensation for 20 particles.

The purpose of the present paper is to extend the applications to
arbitrary scattering lengths and large particle numbers.  We want to
extract the general properties of the solutions especially for large
scattering lengths where mean-field computations are invalid.  We
obtain naturally self-bound many-body systems, even when the two- and
three-body subsystems are unbound.  The paper begins with a brief
description of the hyperspherical adiabatic expansion method in
section~\ref{sec:hypersph-adiab-meth}.  Then in
section~\ref{sec:angular-potentials} the details of the properties of
the angular eigenvalue is discussed as the decisive ingredient in the
radial potential prociding the information about the two-body
interaction.  In section~\ref{sec:radi-potent-solut} we discuss the
radial potential and the properties of the corresponding solutions.
Finally in section~\ref{sec:decay-rates} we discuss stability criteria
for condensates expressed in terms of various time scales and decay
rates.  Section~\ref{sec:conclusion} contains the conclusions.

\section{Hyperspherical adiabatic method}

\label{sec:hypersph-adiab-meth}

We use the hyperspherical adiabatic expansion method with finite-range
two-body interactions and simplifying assumptions about the wave
function.  We shall briefly describe the method and the assumptions.
Details are given in \cite{sor02b}.

The system of $N$ identical particles of mass $m$ is in the center of
mass frame described by hyperspherical coordinates, i.e., one length,
the hyperradius $\rho$, given by
\begin{eqnarray}
  \rho^2&=&\frac1N\sum_{i<j}^Nr_{ij}^2=\sum_{i=1}^Nr_i^2-NR^2
  \;
  \label{eq:rho2}
\end{eqnarray}
and $3N-4$ hyperangles $\Omega$ \cite{bar99a,sor01}. The $i$th
single-particle coordinate is $\vec r_i$, $\vec R$ is the center of
mass coordinate, and
\begin{eqnarray}
  r_{ij}=|\vec r_j-\vec r_i| \equiv
  \sqrt2\rho\sin\alpha_{ij}
  \;,
\end{eqnarray}
with $\alpha_{ij}\in[0,\pi/2]$.  The atoms are trapped in an external
field approximated by a spherically symmetric harmonic oscillator
potential of angular frequency $\omega$:
\begin{eqnarray}
  V\sub{ext}=\sum_{i=1}^N \frac12m\omega^2r_i^2 =  
 \frac12m\omega^2  (\rho^2 + NR^2)  \;.
\end{eqnarray}

Without any two-body interaction between the particles the
ground-state wave function is a Hartree product of Gaussian
amplitudes:
\begin{eqnarray}
  \Psi\sub{total}
  =\prod_{i=1}^N e^{-r_i^2/(2b\sub t^2)}
  =e^{-\rho^2/(2b\sub t^2)} e^{-NR^2/(2b\sub t^2)} \;,
\end{eqnarray}
where the trap length is $b\sub t=\sqrt{\hbar/(m\omega)}$. The second
radial moments are $\langle r_i^2\rangle=3b\sub t^2/2$ and $\langle
R^2\rangle=3b\sub t^2/(2N)$.  For large $N$ the average hyperradius
therefore approaches the average mean-field radial coordinate times
$\sqrt N$, see eq.~(\ref{eq:rho2}).  The hyperangles $\Omega$
determine the relative orientations of the particles.

With a two-body interaction term $V(r_{ij})$ the total Hamiltonian
becomes
\begin{eqnarray}
  \hat H&=&
  \sum_{i=1}^N\Big(\frac{\hat p_i^2}{2m}+\frac12m\omega^2r_i^2\Big)+
  \sum_{i<j}^NV(r_{ij}) \; .
  \label{eq:Hamiltonian}
\end{eqnarray}
It separates into a center of mass part ($ \hat H\sub{c.m.}$), a
radial part ($\hat H_\rho$), and an angular part ($\hat h_\Omega$)
depending respectively on $\vec R$, $\rho$, and $\Omega$
\cite{sor01}:
\begin{eqnarray}
  \label{eq:1}
  \hat H &=&
  \hat H\sub{c.m.}
  +
  \hat H_\rho+\frac{\hbar^2\hat h_\Omega}{2m\rho^2} \;, \\
  \label{eq:2}
  \hat H\sub{c.m.}
  &=&
  \frac{\hat p_R^2}{2Nm}+\frac12Nm\omega^2R^2
  \;,
  \\
  \label{eq:3}
  \hat H_\rho
  &=&
  \hat T_\rho+\frac12m\omega^2\rho^2
  \;,
  \\
  \label{eq:4}
  \frac{\hbar^2\hat h_\Omega}{2m\rho^2}
  &=&
  \hat T_\Omega+\sum_{i<j}V_{ij}
  \;,
\end{eqnarray}
where $\hat T_{\rho}$ and $\hat T_{\Omega}$ are radial and angular
kinetic energy operators.  Then the center of mass motion always
separates from the relative motion since the $V_{ij}$-terms are
independent of $R$.

We remove the center of mass motion and study the Schr\"odinger
equation for relative coordinates
\begin{eqnarray} \label{e10}
  (\hat H-\hat H\sub{c.m.})\Psi = E\Psi
  \;.
\end{eqnarray}
The adiabatic hyperspherical expansion of the wave function is
\begin{eqnarray}
  \Psi(\rho,\Omega)
  &=&
  \rho^{-(3N-4)/2}
  \sum_{\nu=0}^\infty
  f_{\nu}(\rho)\Phi_\nu(\rho,\Omega)
  \;,
  \label{eq:hyperspherical_wave_function}
\end{eqnarray}
where $\Phi_\nu$ is an eigenfunction of the angular part of the
Hamiltonian with an eigenvalue $\hbar^2\lambda_\nu(\rho)/(2m\rho^2)$:
\begin{eqnarray}
  \hat h_\Omega\Phi_\nu(\rho,\Omega)
  &=&
  \lambda_\nu(\rho)\Phi_\nu(\rho,\Omega)
  \;.
  \label{eq:angular_equation}
\end{eqnarray}
Then eq.~(\ref{e10}) leads to a set of coupled radial equations.
Neglecting couplings between the different $\nu$-channels yields the
radial eigenvalue equation:
\begin{eqnarray}
  &&
  \Big(-\frac{\hbar^2}{2m}\frac{d^2}{d\rho^2} + U_\nu(\rho) - E_{\nu}\Big)
  f_{\nu}(\rho)
  =
  0
  \;,
  \label{eq:radial.equation}
  \\
  &&
  \frac{2mU_\nu(\rho)}{\hbar^2}
  =
  \frac{\lambda_\nu}{\rho^2}+
  \frac{(3N-4)(3N-6)}{4\rho^2}+
  \frac{\rho^2}{b\sub t^4}
  \;,
  \label{eq:radial.potential}
\end{eqnarray}
where $E_{\nu}$ is the energy and the adiabatic potential $U_\nu$ acts
as an effective mean-field potential as a function of the
hyperradius. This potential consists of three terms, i.e., the external
field, the generalized centrifugal barrier, and the angular average of
the interactions and kinetic energies.  The neglected non-diagonal
terms are typically about 1\% of the diagonal terms for attractive
Gaussian potentials. 

We have so far no restriction on the many-body wave function, but
include in principle any structure of the system. To choose a
convenient form we follow the philosophy in the
Faddeev-Yakubovski{\u\i} formulations \cite{fad61,yak67}, i.e., the
additive decomposition of the wave function reflects explicitly the
possible asymptotic large-distance behaviour of cluster subsystems. We
expect that two-body correlations are most important and we select the
corresponding terms in the decomposition.  Higher-order correlations
are then essentially neglected.  This procedure assumes a very
different starting point compared to the Jastrow factorization into
products of two-body wave functions \cite{bij40,din49,jas55}. The
traditional Jastrow form is expected to be more efficient for large
densities while our method is well suited for the low densities
encountered for Bose-Einstein condensates.

Emphasizing two-body correlations we therefore decompose the angular
wave function $\Phi$ in the symmetric Faddeev components $\phi$
\begin{eqnarray} \label{e15}
  \Phi(\rho,\Omega) 
  =
  \sum_{i<j}^N  \phi_{ij}(\rho,\Omega) 
  \approx
  \sum_{i<j}^N  \phi(\rho,r_{ij})
  \;,
\end{eqnarray}
where the last approximation assumes that only relative $s$-waves
between each pair of particles contribute. Then the coordinate
dependence reduces to the distance $r_{ij} =
\sqrt2\rho\sin\alpha_{ij}$. Neglecting higher-order partial waves
is justified when the large-distance properties are decisive.  The
capability of this assumption for large scattering length has been
demonstrated for $N=3$ by describing the intricate Efimov effect
\cite{fed93,jen97}.

The angular eigenvalue equation (\ref{eq:angular_equation}) can by a
variational technique be rewritten as a second order
integro-differential equation in the variable $\alpha_{12}$
\cite{sor01}.  For atomic condensates the interaction range is very
short compared to the spatial extension of the $N$-body system. Then
this equation simplifies even further to contain at most
one-dimensional integrals.  The validity of our approximations only
relies on the small \emph{range} of the potential, whereas the
scattering length can be as large as desired.

We shall use the finite-range Gaussian potential $V(r) =
V_0\exp(-r^2/b^2)$.  Thus we have either overall attractive or overall
repulsive potentials depending on the sign of the strength $V_0$.  It
is convenient to measure the strength of the interaction in units of
the Born-approximation of the scattering length
\begin{eqnarray}
  a\sub B \equiv
  \frac{m}{4\pi\hbar^2}\int d^3\vec r_{kl}\; V(\vec r_{kl})
  =
  \frac{\sqrt\pi mb^3V_0}{4\hbar^2}
  \;,\;
  \label{eq:7}
\end{eqnarray}
where the last expression is for the Gaussian potential.  We use the
sign convention that the scattering length $a_s>0$ for a purely
repulsive potential, such that $a_s \simeq a\sub B$ for $|a\sub
B|/b\ll1$.  Thus $a_s>0$ for purely repulsive potentials while purely
attractive potential can lead to any, positive or negative, value of
$a_s$ depending on $V_0$ and $b$.  In
appendix~\ref{sec:numerical-details} is collected the connections
between the Gaussian strength measured in $a\sub B/b$ and the
scattering length $a_s/b$ for the cases applied in this work.  In
most of the numerical work we have $|a\sub B|/b$ close to unity.

\section{Angular potentials}

\label{sec:angular-potentials}

The key quantity in the radial equation~(\ref{eq:radial.equation}) is
the angular eigenvalue $\lambda$ obtained from
eq.~(\ref{eq:angular_equation}).  This eigenvalue depends on the
number of particles, on the size of the system through the
hyperradius, and on the two-body potential through the scattering
length.  The behaviour of $\lambda$ is decisive for the effective
potential in eq.~(\ref{eq:radial.potential}) which in turn determines
the properties of the solutions to eq.~(\ref{eq:radial.equation}).  We
shall therefore first study the dependence of $\lambda$ on the
parameters in the model.  We use the method described in
\cite{sor02b}.  The two-body interaction is a simple Gaussian either
purely attractive or purely repulsive.  This finite-range interaction
never produces the collapse at short distance arising from an
attractive $\delta$-force \cite{fed01}.  Thus we can as well use
attractive potentials with one or more bound states.

\subsection{General eigenvalue behaviour}

The angular eigenvalue spectrum coincides with the free spectrum
(without interaction) at both small and large hyperradii; for $\rho =
0$ because all interactions are multiplied by $\rho^2$, see
eqs.~(\ref{eq:1}) and (\ref{eq:4}), and at $\rho = \infty$ because the
short-range interaction has no effect at large distances.  Thus,
perturbation theory for small $\rho$ for a Gaussian potential shows
that the eigenvalues change from their hyperspherical values
$\lambda_\nu(0) = 2\nu(2\nu+3N-5)$, $\nu=0,1,\ldots$, as
\begin{eqnarray}
  \lambda_\nu(\rho)-\lambda_\nu(0)
  =
  \frac{m V_0}{\hbar^2} N(N-1) \rho^2
  \;.
\end{eqnarray}

If the two-body potential is attractive, but too weak to support any
bound state, the eigenvalues reach a minimum as function of $\rho$ and
then return to one of the finite hyperspherical values.  For more
attractive potentials there is a one-to-one correspondence between one
given two-body bound state of energy $E^{(2)}<0$ and one eigenvalue
$\lambda$ diverging with $\rho$ as $\lambda = 2mE^{(2)}\rho^2 /\hbar^2
$. The corresponding structure describes, appropriately symmetrized,
one pair of particles in that bound state and all others far apart
from the pair and from each other. In addition to this finite number
of such negative eigenvalues the hyperspherical spectrum emerges at
large distances.

\begin{figure}[htb]
  \begin{center}
    \input{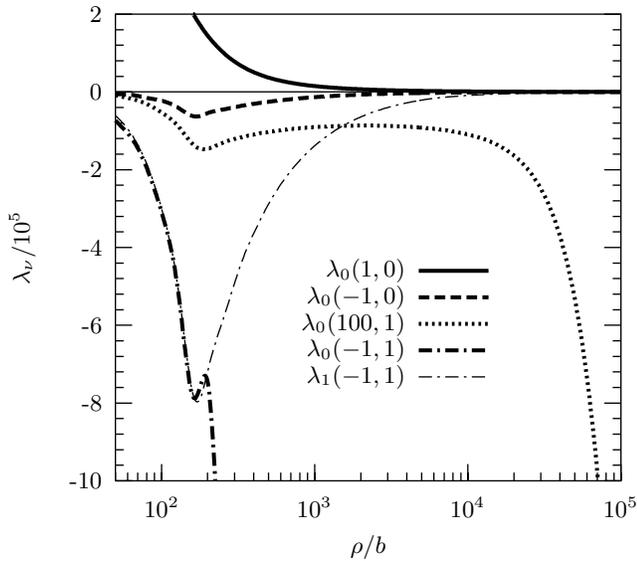}
  \end{center} 
  \caption []
  {Angular eigenvalues $\lambda_\nu$ (numbered with inceasing $\nu$ as
    $\nu=0,1$) as functions of hyperradius divided by interaction
    range, $\rho /b$, for $N=100$, for different scattering lengths
    $a_s/b$ and numbers of bound two-body states $\mathcal N\sub
    B$ indicated as $\lambda_\nu(a_s/b,\mathcal N\sub B)$ on the
    figure.}
  \label{fig:as_var}
\end{figure}

To illustrate we show in fig.~\ref{fig:as_var} a number of possible
angular eigenvalues $\lambda$ as functions of hyperradius for
different potentials.  The entirely positive (solid) curve corresponds
to a repulsive Gaussian.  The diverging (dotted and thick dot-dashed)
curves correspond to potentials with one bound two-body state. For our
purpose the curves approaching zero for large $\rho$ (dashed and thin
dot-dashed curves) are the most interesting, since they are crucial
for the later description of the condensate.  This is true even when
the potential has lower-lying bound states corresponding to diverging
$\lambda$ (thick dot-dashed curve).

The convergence of $\lambda$ as $\rho \rightarrow 0$ is due to the
finite range of the potential and the behaviour depends on the
interaction range $b$.  The deep minima in fig.~\ref{fig:as_var} at
small to intermediate distances depend strongly on both the number of
particles and the strength of the attraction. They are substantially
deeper than reported in \cite{sor02,sor02b} where one term
inadvertently was used with the wrong sign in the numerical examples.
This error is only significant for small and intermediate $\rho$.
Increasing the strength of the attraction always leads to larger
negative values of $\lambda$.  However, at some point one more bound
two-body state reveals its presence by changing convergence to zero
into a parabolic divergence with $\rho$.

For distances much larger than the range of the potential the
eigenvalues could as well be computed from a zero-range interaction,
i.e., $4\pi\hbar^2a_s\delta(\vec r)/m$. The hyperharmonic angular wave
function should then be appropriate for $\Phi$ and the eigenvalue
$\lambda$ obtained as the corresponding expectation value. The lowest
hyperharmonic is a constant independent of angles and the result is
\cite{boh98}
\begin{eqnarray}
  \lambda_\delta(N,\rho)
  &=&
  \sqrt{\frac{2}{\pi}}
  \;
  \gammafktB{3N-3}{3N-6}
  \; N (N-1) \; \frac{a_s}{\rho}
  \nonumber\\
  &\stackrel{N\gg1}{\longrightarrow}&
  \frac{3}{2}\sqrt{\frac{3}{\pi}}N^{7/2}\frac{a_s}{\rho}
  \label{eq:lambda_delta}
  \;.
\end{eqnarray}
This zero-range result is inversely proportional to $\rho$ for all
hyperradii and consequently with a non-physical divergence when $\rho
\rightarrow 0$.  The only length scale arises from the strength of the
$\delta$-function.  In mean-field calculations this strength is chosen
to reproduce the correct scattering length $a\sub B$ in the Born
approximation \cite{pet01,esr99b}.  To reach this limit with a
Gaussian potential then requires that the $\delta$-function is
approached while $a\sub B$ is maintained equal to the desired value of
$a_s$.

This artificial construction is due to the lack of correlations in
mean-field computations where the effective interaction is adjusted to
the available Hilbert space. We use finite-range Gaussian potentials
and include two-body correlations. Then we expect the large-distance
asymptotic behaviour to be described by eq.~(\ref{eq:lambda_delta})
with the correct scattering length.  This tests the efficiency of the
simplified structure of the wave function in eq.~(\ref{e15}).
Mathematically this should result from the structure of the second
order integro-differential angular eigenvalue equation
\cite{sor01,sor02b}.

\begin{figure}[htb]
  \begin{center}
    \input{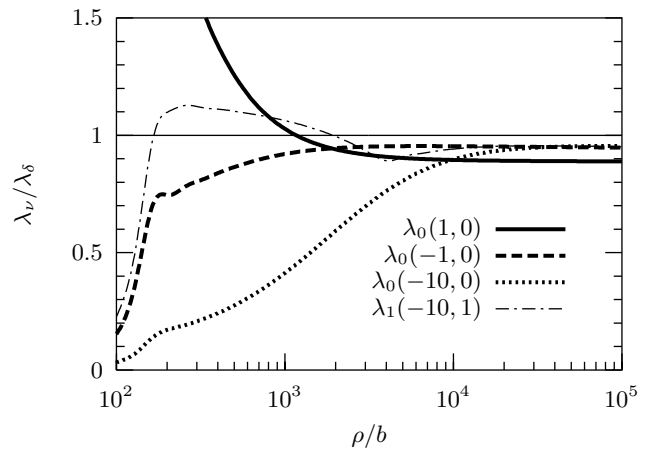}
  \end{center}
  \caption
  [] {Same as figure \ref{fig:as_var}, but the angular potential is
    shown in units of the zero-range result in
    eq.~(\ref{eq:lambda_delta}) as obtained in \cite{boh98}.}
  \label{fig:as_var.transformed}
\end{figure}

Numerically we investigate the asymptotic behaviour of $\lambda$ in
this context by comparing to the zero-range result $\lambda_\delta$ in
fig.~\ref{fig:as_var.transformed}.  The convergence to the limiting
value is fastest for the smallest value of $|a_s|$ (dashed and
solid curve) reflecting that the correlations arising for large
scattering lengths (dotted line) cannot be accounted for by the
zero-range result. This is well understood for three particles where
the Efimov effect (very large $a_s$) extends correlations in
hyperradius to distances around four times the average scattering
length \cite{fed93,jen97}.  These effects are not present in the
zero-range expectation value contained in $\lambda_\delta$.  When
$\rho$ exceeds $|a_s|$ by a sufficiently large amount the Efimov
effect disappears in $\lambda$ and $\lambda_\delta$ is approached.

A stronger attraction corresponding to one two-body bound state
produces one diverging eigenvalue (figure~\ref{fig:as_var}) while the
second eigenvalue converges towards $\lambda_\delta$ (dot-dashed
curve).  In fact $\lambda_1(-10,1)$ almost coincides with the lowest
eigenvalue $\lambda_0(-10,0)$ for the same scattering length but for a
potential without bound two-body states (dotted curve).

The numerical deviations from $\lambda_\delta$ at large distance is in
all cases less than 10\%. The asymptotic behaviour is very smooth but
still originating in systematic numerical inaccuracies.

These results demonstrate that the scattering length entirely
determines the asymptotic behaviour of the potentials.  The radial
shape of the two-body potential could be Gaussian, square-well,
Woods-Saxon, or Yukawa, still the same $a_s$ would produce the
same angular eigenvalue at sufficiently large distance.

\subsection{$N$-dependence}

\begin{figure}[htb]
  \input{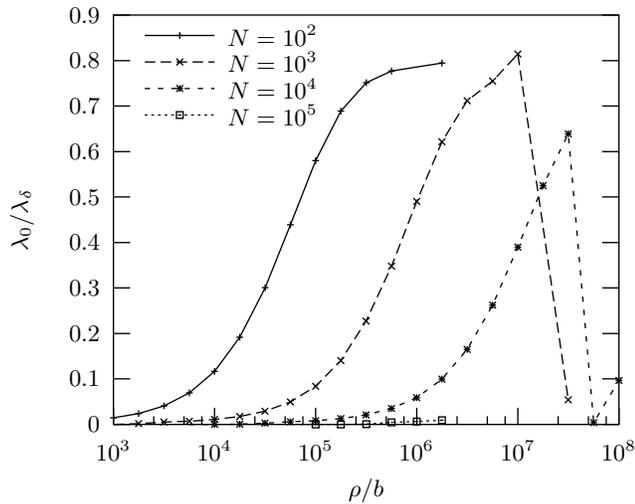}
  \caption[] {The lowest angular eigenvalue in
    units of $\lambda_\delta$ as a function of hyperradius for $a\sub
    s/b=-401$ for four different numbers of particles
    $N=10^2,10^3,10^4,10^5$.}
  \label{fig:lambda.401}
\end{figure}

The angular eigenvalues increase rapidly with $N$ as seen already from
the $N^{7/2}$-dependence in $\lambda_\delta$. The major variation in
magnitude is then accounted for by using this large-distance
zero-range result as the scaling unit. We show in
fig.~\ref{fig:lambda.401} a series of calculations for the same
two-body interaction for different numbers of atoms.  All curves are
similar with a systematic increase in the characteristic hyperradius
$\rho_a$ where they bend over and approach the zero-range result.  We
then numerically determine this characteristic length $\rho_a$ to be
proportional to the scattering length and a particular power $7/6$ of
$N$, i.e.,
\begin{eqnarray}
  \rho_a(N)\equiv|a_s|N^{7/6} 
  \;.
  \label{eq:rhoa}
\end{eqnarray}

\begin{figure}[htb]
  \input{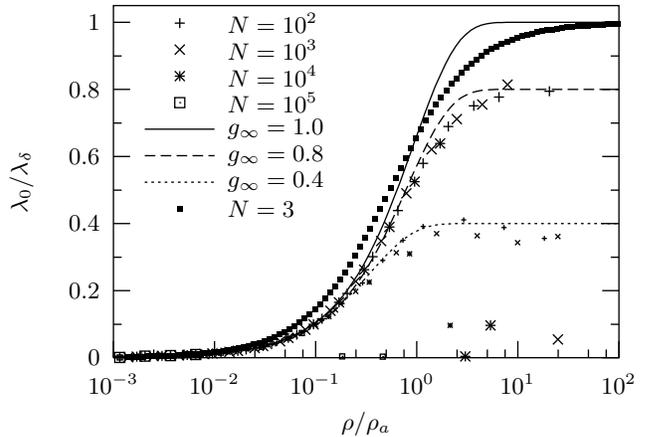}
  \caption[] {The same as fig.~\ref{fig:lambda.401}, but with $\rho$
    in units of $\rho_a$.  The points following the intermediate curve
    ($g_\infty=0.8$) are obtained with many integration points and the
    points along the lower curve ($g_\infty=0.4$) are obtained with
    fewer points. The curve for $g_\infty=1.0$ is the expected correct
    asymptotic behaviour from eq.~(\ref{e19}).  The points for $N=3$
    are calculated with the zero-range model from \cite{fed01b}.}
  \label{fig:lambda.401.transformed}
\end{figure}

The quality of this scaling is illustrated in
fig.~\ref{fig:lambda.401.transformed}, where all curves essentially
coincide for distances smaller than $\rho_a$.  At larger hyperradii
the zero-range result of $+1$ should be obtained.  However, systematic
deviations from a common curve is apparent.  For each $N$ one smooth
curve is followed at small and intermediate distances implying that
the numerical inaccuracies here are systematic until random
fluctuations set in at large $\rho$.

The smooth numerical curves can be rather well reproduced by the
function
\begin{eqnarray} \label{e19}
  \lambda^{(-)}(N,\rho)
  =
  \lambda_\delta(N,\rho)\cdot g^{(-)}(\rho/\rho_a)
  \;,
  \\ \label{e20}
  g^{(-)}(x)
  =
  g_{\infty} \big(1-e^{-x/x_a}\big)\Big(1+\frac{x_b}{x}\Big)
  \;,
\end{eqnarray}
where $g_{\infty}=1$ in accurate calculations.  The exponential term
is introduced to reproduce the rather fast approach to the asymptotic
value as seen in fig.~\ref{fig:lambda.401.transformed}. The behaviour
at smaller distance, depending on the range of the interaction, is
simulated by the $x_b$-term.  The extreme limit of $\rho \rightarrow
0$ is attempted reproduced on the function in eq.~(\ref{e20}).

\begin{table}[htb]
  \caption[] {Numerical values of $g_{\infty}$, $x_a$, and $x_b$ for four
    scattering lengths.}  
  \begin{tabular}{|c||c|c|c|c|}
    \hline    
    $a_s/b$ & $-5.98$ & $-401$ & $-799$ & $-4212$ \\
    \hline
    $g_{\infty}$ & 0.99 & 0.80 & 0.65 & 0.30  \\
    \hline
    $x_a$ & 1.06 & 0.74 & 0.59 & 0.28  \\
    \hline
    $g_{\infty}/x_a$ & 0.93 & 1.081 & 1.099 & 1.077  \\
    \hline
    $x_b$ & $0.15$ & $2.3\cdot10^{-3}$ & $1.15\cdot10^{-3}$ & 
    $2.2\cdot10^{-4}$  \\
    \hline
    $x_b/(b/|a_s|)$ & 0.92 & 0.922 & 0.919 & 0.927  \\
    \hline
  \end{tabular}
  \label{table:parameters}
\end{table}
The two groups of computations in
fig.~\ref{fig:lambda.401.transformed} are reasonably well reproduced
by the parameter sets $x_a \simeq 0.74$, $x_b \simeq 2.3\cdot10^{-3}$,
and $ g_{\infty} \simeq 0.8$ or $g_{\infty} \simeq 0.4$. These
parameters may depend on the scattering length, and we therefore
repeated the computation for various $a_s$.  The best choice of
parameters are shown in table~\ref{table:parameters}.  We notice that
$g_{\infty}$ and $x_a$ both are of order unity, and that the fraction
$g_{\infty}/x_a$ is almost constant, except for the smallest
scattering length.  The parameter $x_b$, introduced to account for the
finite interaction range, is almost equal to $b/|a\sub s|$.  At large
hyperradii, where $\rho \gg \rho_a$, $\lambda^{(-)}$ approaches
$g_{\infty} \lambda_\delta$.  The rather accurate results for $N=100$
displayed in fig.~\ref{fig:as_var.transformed} confirm that
$g_\infty\simeq1$ by deviating less than $10\%$ from $\lambda_\delta$
at large hyperradii.

The angular eigenvalue is given by $g^{(-)}(x) \simeq g_\infty x/x_a$
for $x_b\ll \rho/\rho_a \ll x_a$.  Numerical calculations in this
intermediate region of hyperradii therefore rather accurately
determines the fraction $g_{\infty}/x_a \simeq 1.08 $ as given in
table~\ref{table:parameters}. With $g_{\infty}=1$ this implies that
$x_a\simeq1/1.08\simeq0.92$. The parameters of $g^{(-)}(x)$ in
eq.~(\ref{e20}) are then given by
\begin{eqnarray} \label{e22}
  g_{\infty}=1
  \;,\quad
  x_a \simeq 0.92
  \;,\quad
  x_b \simeq 0.92 \frac{b}{|a_s|}
  \;.
\end{eqnarray}

We can compare with the rigorous result for $N=3$ \cite{jen97} where
the angular eigenvalue at large-distance coincide with
$\lambda_\delta$ in eq.~(\ref{eq:lambda_delta}), i.e.,
\begin{eqnarray}
  \lambda_\delta(N=3,\rho) =
  \frac{48a_s}{\sqrt2\pi\rho}
  \;.
\end{eqnarray}
Thus also for $N=3$ the universal function $g^{(-)}$ asymptotically
approaches $1$ for all scattering lengths.  More accurate results for
$N=3$ have been calculated with the zero-range model from
\cite{fed01b} and are shown in rescaled form in
fig.~\ref{fig:lambda.401.transformed}.  The behaviour is similar to
the behaviour of the eigenvalue for the $N$-body systems, which
confirms the schematic model.

\begin{figure}[htb]
  \input{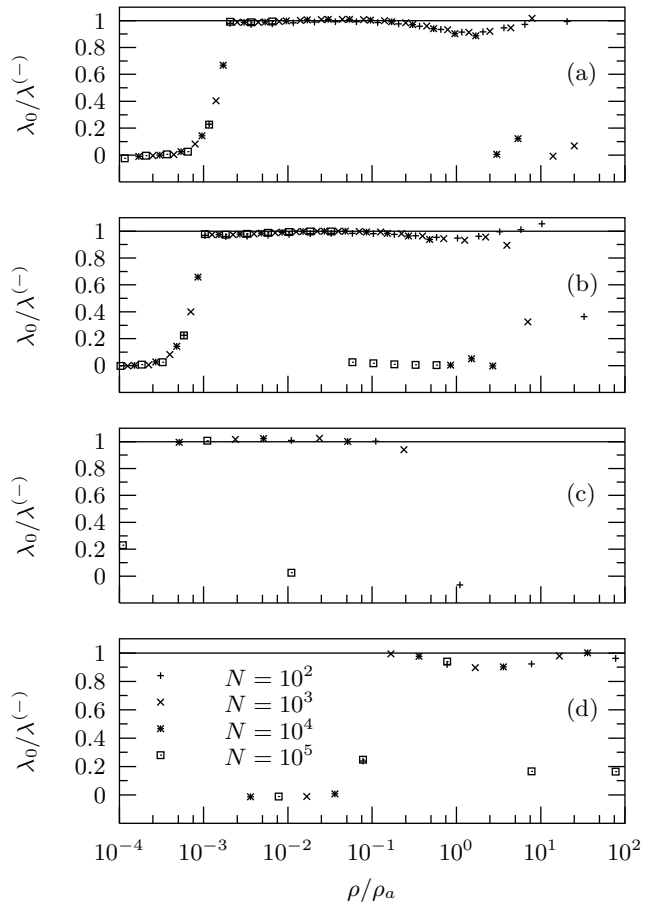}
  \caption[] {The lowest angular eigenvalue $\lambda_0$ in units of 
    $\lambda^{(-)}$, eqs.~(\ref{e19}) and (\ref{e20}) and
    table~\ref{table:parameters}, as functions of the hyperradius in
    units of $\rho_a$, eq.~(\ref{eq:rhoa}).  The scattering lengths
    are given by a) $a_s/b=-401$, b) $a_s/b=-799$, c) $a\sub
    s/b=-4212$, and d) $a_s/b=-5.98$. The different $N$-values are
    as indicated.}
  \label{fig:lambda_analytic}
\end{figure}

The accuracy of the parametrization in eq.~(\ref{e20}) is seen in
figs.~\ref{fig:lambda_analytic}a-d, where the angular eigenvalues are
shown in units of $\lambda^{(-)}$ with the individual set of
parameters from table~\ref{table:parameters}.  Good agreement is found
for $\rho/\rho_a>x_b$, except at large hyperradii where the numerical
inaccuracy increases with increasing scattering length.  Fortunately,
the large-distance behaviour is known from analytic considerations and
we do not need to rely on numerical computations at these distances.
The remaining deviations occur at small hyperradii for
$\rho/\rho_a<x_b$ or equivalently for $\rho<N^{7/6}b$, where the
result depends on the radial shape of the two-body interaction.

\subsection{$N$-dependence with bound two-body states} 

In the presence of a bound two-body state of energy $E^{(2)}$ one
angular eigenvalue eventually diverges at large hyperradii as
\cite{nie01}
\begin{eqnarray}
  \label{e24}
  \lambda^{(2)}(\rho)=\frac{2m\rho^2}{\hbar^2}E^{(2)}
  \;,\qquad
  E^{(2)}<0
  \;.
\end{eqnarray}
In the limit of weak binding, or for numerically large scattering
lengths, the energy of the two-body bound or virtual state is given by
\begin{eqnarray} 
  E^{(2)}=-\frac{\hbar^2}{m a_s^2}c
  \;,
  \label{e25}
\end{eqnarray}
where $c$ approaches unity for large scattering lengths.

We now parametrize the angular eigenvalue by an expression similar to
eqs.~(\ref{e19}) and (\ref{e20}). The effect of the bound two-body
state is only expected to show up at large distances where the
behaviour corresponds to eq.~(\ref{e25}). The small and intermediate
distances resemble the behaviour when no bound state is present.
Therefore we arrive at the parametrization
\begin{eqnarray} \label{e26}
  &&
  \lambda^{(+)}(N,\rho)
  =
  \lambda_\delta(N,\rho) \; g^{(+)}(\rho/\rho_a)
  \;,
  \\  \label{e27}
  &&
  g^{(+)}(x)
  =
  -x
  \Big(1+\frac{x_b}{x}\Big)
  \bigg(\frac{g_{\infty}}{x_a} + c\frac{4}{3}\sqrt{\frac{\pi}{3}}x^2\bigg)
  \;,
\end{eqnarray}
with the notation and estimates from eq.~(\ref{e22}).

\begin{figure}[htb]
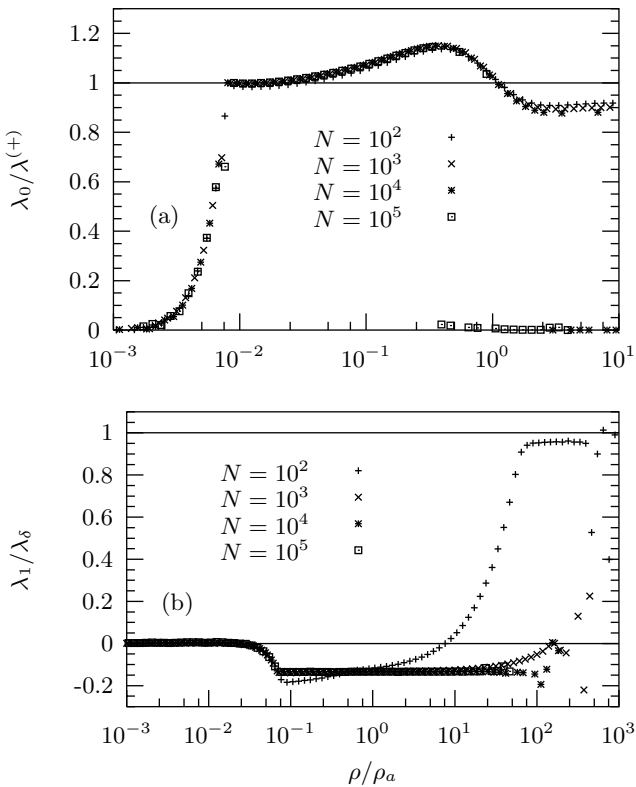

  \input{olep2fig6a.tex}
  \input{olep2fig6b.tex}
  \caption[] {a) The lowest angular eigenvalue $\lambda_0$ in units of
    $\lambda^{(+)}$, eqs.~(\ref{e26}) and (\ref{e27}), for $a\sub
    s/b=+100$ and $c=1.02$, when the potential holds one bound
    two-body state. The number of particles is indicated on the
    figure. The parameters are $g_{\infty}/x_a=-1.09$ and
    $x_b=9.2\cdot10^{-3}$.  b) The first excited angular eigenvalue
    $\lambda_1$ in units of $\lambda_\delta$ for $a_s/b = +10$.}
  \label{fig:lambda_analytic.plus100}
\end{figure}

We compare in fig.~\ref{fig:lambda_analytic.plus100} the
parametrization in eqs.~(\ref{e26}) and (\ref{e27}) with the computed
angular eigenvalues for a potential with one bound two-body state.
For the large scattering length ($a_s/b=100$) in
fig.~\ref{fig:lambda_analytic.plus100}a one smooth curve applies for
all the particle numbers; numerical inaccuracies set in at larger
hyperradii, which is most obvious for the largest particle numbers.
This smooth curve is in a large interval of hyperradii at most
deviating by 20\% from the parametrized form, and even less than 10\%
at large hyperradii, before the numerical instability sets in.

For smaller $a_s$ ($a_s/b=+10$) the deviation at large
hyperradii is less than $1\%$.  The deviation at intermediate
distances would decrease by inclusion of a linear term in
eq.~(\ref{e27}).  The smooth curve at small hyperradii is outside the
range of validity of the parametrization, i.e., within the range of
the two-body potential and then depending on details of the
interaction.

The lowest eigenvalue $\lambda_0$ diverges at large hyperradius as
described by eq.~(\ref{e26}).  If the two-body potential only has one
bound state the second eigenvalue $\lambda_1$ is expected to approach
zero at large distances as $\lambda_\delta$. This pattern should be
repeated with more than one bound two-body state, i.e., the first
non-divergent angular eigenvalue would behave as $\lambda_\delta$ for
large $\rho$.

We therefore in fig.~\ref{fig:lambda_analytic.plus100}b compare the
computed first excited angular eigenvalue with $\lambda_\delta$ for
different $N$. As in fig.~\ref{fig:lambda.401.transformed} we obtain
smooth and almost universal curves at large $\rho$, where the approach
to unity sets in exponentially fast depending on $N$, but now much
later when $\rho \sim 10^2 \rho_a$.  Clearly a parametrization would
also here be possible.

The large-distance asymptotic behaviour of $\lambda_1$ now corresponds
to an effectively repulsive potential.  However, at small and
intermediate hyperradii the potential is still effectively attractive
($\lambda_1<0$).  This attractive region may support a self-bound
system located at distances far inside and independent of the
confining external field.

This feature is absent in the mean-field description of Bose-Einstein
condensation.  For overall repulsive potentials corresponding to
positive scattering lengths no attractive part is possible.  For
attractive potentials either a zero-range potential would produce a
collapsed wave function and a finite-range potential would not give
repulsion at large distance.

\subsection{The properties of $\lambda^{(\pm)}$}

The functions $\lambda^{(\pm)}$ coincide when $\rho\ll \rho_a$ and
depends only on the absolute magnitude of the scattering length
$|a_s|$.  For $\rho\gg \rho_a$ the functions differ qualitatively,
i.e., $\lambda^{(-)}$ approaches zero as $\lambda_\delta$ while
$\lambda^{(+)}$ diverges as $-\rho^2$.

At intermediate hyperradii, $x_b\ll \rho/\rho_a \ll x_a$, when
\begin{eqnarray}
  b \ll \frac{\rho}{N^{7/6}} \ll |a_s|
  \;,
  \label{eq:intermediate_region}
\end{eqnarray}
the angular eigenvalue $\lambda^{(\pm)}$ approaches a constant value
$\lambda_\infty$, i.e.,
\begin{eqnarray}
  \label{e28}
  \lambda_\infty
  \equiv
  \lambda_\delta \frac{\rho g_{\infty}}{\rho_a x_a}
  = - \frac{3g_{\infty}}{2x_a} N^{7/3} \sqrt{\frac{3}{\pi}} 
  \simeq -1.59N^{7/3}  
  \;.
\end{eqnarray}
This numerical result is in agreement with the following derivation.

The angular eigenvalue for large scattering length $a_s$ is
independent of hyperradius $\rho$ when $\rho$ is large compared to the
range $b$ of the potential but small compared to $a_s$.  The plateau
value $\lambda_\infty$ can be estimated as the intersection between
two curves at the point $\rho_a$.  The first curve is the
parabollically decreasing $\lambda(\rho)$ corresponding to a bound
two-body state, i.e., $\lambda(\rho)=2m\rho^2E^{(2)}/\hbar^2 = -
2\rho^2/a_s^2$, where $E^{(2)}$ is given by eq.~(\ref{e25}) with
$c=1$.  The second curve is the increasing $\lambda_\delta(\rho)$ for
an attractive potential ($a_s<0$), see eq.~(\ref{eq:lambda_delta}).
Thus $\lambda_\delta(\rho_a)=\lambda(\rho_a)$ gives
\begin{eqnarray}
  \rho_a &\simeq& \sqrt[3]{\frac34}N^{7/6}|a_s| \;,
  \\
  \lambda_\infty(N) &\simeq& -\sqrt[3]{\frac{9}{2}}N^{7/3} \simeq
  -1.65N^{7/3} \;,
\end{eqnarray}
which is very close to the numerical results in eqs.~(\ref{eq:rhoa})
and (\ref{e28}).

The symbol $\lambda_\infty$ is chosen for this constant, since the
$\rho$-region where $\lambda=\lambda_\infty$ increases proportional to
$|a_s|$, see eq.~(\ref{eq:intermediate_region}), and thus extends to
infinity for $|a_s|=\infty$.  With no bound two-body states ($a\sub
s<0$) the lowest angular eigenvalue approaches zero at larger
hyperradii, whereas it diverges towards $-\infty$ as $\rho^2$ when a
bound two-body state is present ($a_s>0$).  On the threshold for a
two-body bound state $a_s= \pm \infty$ and the angular eigenvalue
therefore remains constant. 

In \cite{sor02} $\lambda_\infty \simeq -5N^2$ was estimated by
courageous extrapolation of calculations for $N=10,20,30$ and the
analytic result for $N=3$. The much better estimate in eq.~(\ref{e28})
of the large-$N$ asymptotics of $\lambda_\infty(N)$ increases with a
slightly higher power of $N$ but with a smaller proportionality
factor.

\section{Radial potentials and solutions}

\label{sec:radi-potent-solut}

The radial equation is the next step in the process of obtaining
knowledge about the physical properties of the many-boson system.  The
angular potentials found in the previous section now enter the
effective radial potential in eq.~(\ref{eq:radial.potential}) and
infer information about the interactions to quantities like energy,
size, and structure of the system.

\subsection{Properties of the radial potential}

\begin{figure}[htb]
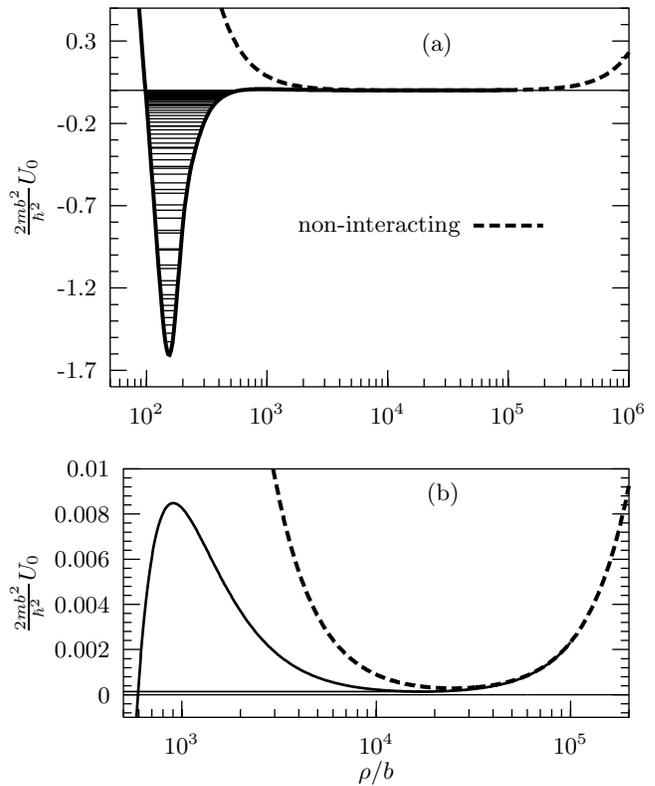

  \begin{center}
    \input{olep2fig7a.tex}
    \input{olep2fig7b.tex}
  \end{center}
  \caption
  [Radial potential at small hyperradii] {a) Radial potential $U_0$
    from eq.~(\ref{eq:radial.potential}) corresponding to the lowest
    angular potentials for $N=100$ and $a_s/b = -1.0$.  We model
    the experimentally studied systems \cite{rob01} of $^{85}$Rb-atoms
    with oscillator frequency $\nu=\omega/(2\pi)=205$ Hz and
    interaction range $b=10$ a.u., thus yielding $b\sub
    t\equiv\sqrt{\hbar/(m\omega)}=1442b$.  Also shown as horisontal
    lines are the negative energies $E_{0,n}$, $n=0,\ldots,58$ in the
    lowest potential in the uncoupled radial equation,
    eq.~(\ref{eq:radial.equation}).  b) Detail at larger hyperradii.
    The energy of the first oscillator-like state (see text) is shown
    as a horisontal line close to zero.}
  \label{fig:u_asneg}
\end{figure}

The radial potential in eq.~(\ref{eq:radial.potential}) consists of
three terms where the repulsive centrifugal barrier and the confining
external field both are positive. The interaction term $\lambda_\nu$
can be either repulsive or attractive depending on hyperradius and
which eigenvalue we consider.  The combination has structure depending
on the interaction. For a purely vanishing or repulsive two-body
potential we arrive at a simple behaviour qualitatively similar to the
non-interacting (dashed) curve shown in fig.~\ref{fig:u_asneg} for
$N=100$.  All solutions are confined to the region between the
infinitely large potential walls at small and large hyperradii.

For a moderately attractive two-body potential a different structure
already appears for the lowest angular potential (solid curve in
fig.~\ref{fig:u_asneg}).  The large-distance behaviour is determined
by the trap and is roughly as without interaction, but the barrier at
intermediate distance is now finite both in height and width.  The
barrier height is small compared to the potential at both small and
large hyperradii.  At smaller hyperradii a rather deep and relatively
narrow minimum is present outside a hard core repulsion.  The minimum
occurs for $N=100$ at about 150 times the range of the interaction
which corresponds to a mean distance $(2\langle\rho^2\rangle/N)^{1/2}$
between each pair of particles of about 15 times the interaction range
$b$.

With this potential we solve the diagonal radial equation.  The
solutions can be divided into groups related to either the first or
the second minimum.  The lowest-lying of the first group of solutions
have negative energies.  In the model they are truly bound states as
they cannot decay into continuum states at large hyperradii
\cite{sor02}.  Their properties are independent of the external trap
which only has an influence at much larger distances. These self-bound
$N$-body states can decay into lower-lying states consisting of
various bound cluster states, e.g., diatomic or triatomic clusters.
The possibility of self-bound many-body systems, even though the two-
and three-body sub-systems are unbound, is also discussed by Bulgac
\cite{bul02}, who, however, considers the three-body interaction
strength as a determining parameter for the properties of the
self-bound many-boson system.

The group of states in the higher-lying minimum at larger distance all
have positive energies.  They are only stable due to the confining
effect of the external trap potential. The lowest of these is
interpreted as the state of the condensate and indicated by a
horizontal line in fig.~\ref{fig:u_asneg}b.  This second minimum
almost coincides with the minimum of the radial potential arising
without any two-body interaction. Thus the structure of the condensate
is similar for both positive and negative scattering lengths arising
from either attractive or repulsive interactions.  However, an
attraction produces in addition a series of lower-lying states at
smaller hyperradii.

Increasing $N$ leaves semi-quantitatively the same features for pure
repulsion, whereas an unchanged attraction leads to decreasing
barriers at intermediate hyperradius and at some point this barrier
vanishes altogether.  At the same time the attractive minimum at
smaller hyperradius becomes deeper.  This in turn leads to an
increasing number of bound states in this minimum as function of $N$.

As the scattering length increases, the barrier disappears and the
effective potential inside the trap has the $\rho^{-2}$-behaviour
characteristic for Efimov states, see eq.~(\ref{eq:radial.potential})
with $\lambda=\lambda_\infty$ of eq.~(\ref{e28}).  The lowest-lying
states are influenced by the details of the two-body interaction and
without Efimov features. However, the higher-lying states, located for
$\rho$-values obeying eq.~(\ref{eq:intermediate_region}), exhibit the
Efimov scaling.  They easily become very large and located far outside
the minimum responsible for the binding.  Only a finite number of
bound states is possible due to the confining external field.

These states are many-body Efimov states arising when the two-body
scattering length is large.  This is also precisely the condition for
the three-body Efimov states \cite{efi70,fed93}.  Therefore the
many-body Efimov states are embedded in the continua of dimer, trimer
and higher-order cluster states.  They could be artifacts of the model
where only special degrees of freedom are treated.  However, these
states may also be distinguishable resonance structures which are
relatively stable because the particles are very far from each other
and the couplings to the continuum states therefore are very weak.  So
far this remains an open question.

\subsection{Interaction energy}

The total energy of a state in the first minimum are independent of
the external field as these states are located at small distances.
These states have no analogue in mean-field calculations.  In
contrast, total energies of the states in the second minimum are
dominated by the contribution from the confining field and therefore
are rather insensitive to anything else than this field and the
corresponding harmonic oscillator quantum numbers.  It is then much
more informative to compare the interaction energies where the large
external field contribution is removed.

\begin{figure}[htb]
  \input{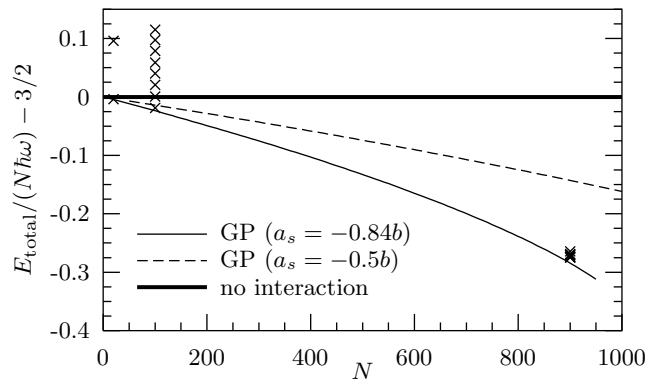}
  \caption[]
  {Gross-Pitaevskii interaction energy as a function of $N$ for $b\sub
    t/b=1442$ for two $a_s$-values.  The points are results from the
    present work for $N=20,100,900$ and $a_s/b=-0.84$ ($a\sub
    B/b=-0.5$).  Only states in the second minimum are displayed.}
  \label{fig:energy_N}
\end{figure}

In fig.~\ref{fig:energy_N} is shown the interaction energy per
particle as a function of the particle number for a relatively weak
attraction corresponding to a small scattering length.  The
Gross-Pitaevskii solution exists and the related interaction energy is
negative due to the attraction between the particles.  A nearly linear
behaviour is observed at small particle numbers, since each particle
interacts with the $N-1$ other particles.  As $N$ increases the
mean-field attraction increases and the Gross-Pitaevskii solution
becomes unstable when $N|a_s|/b\sub t>0.58$.  This corresponds to
$N=1000$ with the present parameters of $|a_s|/b\sub t = 58\cdot
10^{-5}$.

These mean-field interaction energies are in fig.~\ref{fig:energy_N}
compared to the results obtained with the correlated wave functions
from the present formulation.  Only a few of the large number of bound
states for each $N$-value resemble the Gross-Pitaevskii solutions with
a radius corresponding to the second minimum.  For $N=20$ the lowest
six states are located in the first minimum.  Their interaction
energies are large and negative outside the scale of
fig.~\ref{fig:energy_N}.  The seventh state is located in the second
minimum with an interaction energy very close to the Gross-Pitaevskii
result, while the eighth has a positive interaction energy.  This
feature is repeated for increasing $N$, i.e., the lowest state located
in the second minimum is similar to the mean-field result and the
higher-lying states in this second minimum are less bound.  When the
mean-field solutions collapse, the correlated solutions remain stable
due to the use of a finite-range potential.

The correlated and mean-field interaction energies are remarkably
similar when both exist. It may at first appear odd that the
mean-field interaction energy is marginally lower than by use of the
better suited form of the correlated wave function. The reason is that
we compare the mean-field result for an effective interaction which
has the correct scattering length in the Born approximation while the
correlated solution is obtained for an interaction with the correct
scattering length. The mean-field interaction is more attractive to
compensate for the limited mean-field Hilbert space.  The more
revealing comparison is to use the same interaction in both
calculations.

We can then compare results for the same $a\sub B/b=-0.5$, i.e., a
Gross-Pitaevskii calculation with $a_s/b=-0.5$ and a Gaussian of
$a\sub B/b=-0.5$ corresponding to $a_s/b=-0.84$.  As seen in
fig.~\ref{fig:energy_N} (dashed curve) now the mean-field energies are
much smaller.  However, it is remarkable that the correlated solution
essentially reproduces the energy of the mean-field calculation where
the interaction is renormalized to reproduce the correct energy, but
with the wrong wave function.  The implication is that the correlated
wave function is sufficient to describe the correct structure with the
correct interaction.  The large-distance average properties are at
best obtained in mean-field computations, but all features of
correlations are absent by definition.

\subsection{Definition and size of a condensate}

\label{sec:defin-cond}

The lowest-lying positive energy solutions located in the second
minimum have properties similar to the condensates obtained in
Gross-Pitaevskii calculations.  The present formulation also provides
lower-lying negative-energy states.  It is therefore necessary to
discuss how to distinguish a condensate state from other (perhaps very
unstable) $N$-body states.

In mean-field treatments, with repulsive two-body potentials and
confining trap potentials, the condensate is uniquely defined as a
statistical mixture of single-particle states with the ground state
dominating \cite{pet01,pou02}.  A condensate has on average many
particles in the lowest single-particle state.  This many-body state
is only stable against total fragmentation due to the trap and as such
first of all determined by the properties of the trap.  Even with the
trap the many-body state is still at best only approximately
stationary due to the neglected degrees of freedom which allow
energetically favored \mbox{di-,} tri-, and multi-atomic cluster
states.  This instability is also an experimental fact seen by
permanent loss of trapped atoms, e.g., in recombination processes
\cite{don01}.

Without any two-body interaction the properties of the many-body
system is determined by the thick, dashed potential curve in
fig.~\ref{fig:u_asneg}.  Then we can easily identify the condensate as
a state in this potential where the dominating component for finite
temperature is the ground state.  Including attractive two-body
interactions (full curve) the deep minimum at small hyperradius is
produced.  Then the corresponding ground state, located in this
minimum, has nothing to do with a condensate.  The density is so high
that couplings to other degrees of freedom would develop higher-order
correlations and processes like three-body recombinations would
quickly destroy the single-atom nature of the gas.  This $N$-body
ground state does not show the signature of a Bose-Einstein
condensate, where many particles occupy one single-particle level.

The formulation in the present work does not use the concept of
single-particle levels.  Therefore we cannot talk about a statistical
distribution of particles with the majority in the lowest state.
However, we can talk about a many-particle system described as a
superposition of many-body eigenstates, where the lowest states are
favored in thermal equilibrium.  To clarify we can think of a quantum
state $\Psi$ as a superposition of different eigenstates
$\Psi_n(\rho,\Omega)$ in eq.~(\ref{eq:hyperspherical_wave_function})
given by
\begin{eqnarray}
  &&
  \Psi(\rho,\Omega)
  =
  \sum_{n=0}^\infty c_n \Psi_n(\rho,\Omega)
  \nonumber\\
  &&=
  \rho^{-(3N-4)/2}
  \sum_{n=0}^\infty
  c_n
  \sum_{\nu=0}^\infty
  f_{\nu,n}(\rho)\Phi_\nu(\rho,\Omega)
  \;,\quad
\end{eqnarray}
with the normalization $\sum_n|c_n|^2=1$.  A condensate must be
sufficiently large to exceed a certain minimum interparticle distance,
$d\sub c$, below which the atoms are too close and recombine very
fast. This distance depends on the scattering length and on the number
of particles.  Therefore, in our formulation the stationary states
cannot be characterized as a condensate if $\langle r_{ij}^2\rangle
\ll d\sub c^2$ for this wave function.

We define one of the stationary states in this model as the ``ideal
condensate'' state, i.e., the state of lowest energy with
\begin{eqnarray}
  \langle r_{ij}^2\rangle_{n\sub c}
  \gtrsim
  d\sub c
  \;,
\end{eqnarray}
characterized by $n=n\sub c$.  This state is dominated by the
component in the lowest adabatic potential although not necessarily
the states of lowest energy, which might have an average particle
distance less than $d\sub c$.  The appropriate of these excited states
depends on the number of particles and on the scattering length.  The
ideal condensate is then characterized by one dominating component,
with $|c_{n\sub c}| \simeq 1$ and $|c_{n\ne n\sub c}| \ll 1$.

If $d\sub c$ is significantly smaller than the trap length $b\sub t$,
then the state of lowest energy located in the second minimum can be
identified as the condensate.  This state is characterized by a radial
wave function $f(\rho)$ with the root mean square radius
$\langle\rho^2\rangle$ approximately equal to the hyperradius at the
second minimum of the lowest adiabatic potential $U_0(\rho)$.

\begin{figure}[htb]
  \begin{center}
    \input{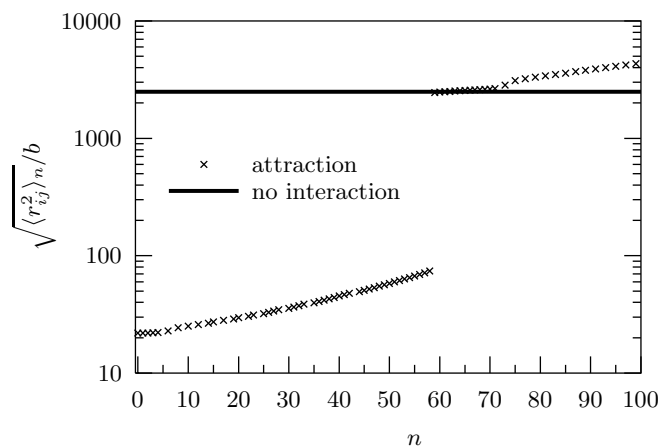}
  \end{center}
  \caption [Interparticle distance as a function of quantum number]
  {The root mean square distance for $\nu=0$ as a function of the
    quantum number for $N=100$, $a_s/b = -1$, $b\sub t/b=1442$.}
  \label{fig:rho2_n.N100}
\end{figure}

To be specific we show in fig.~\ref{fig:rho2_n.N100} the root mean
square interparticle distance given by $\langle r_{ij}^2\rangle_{n} =
2\langle\rho^2\rangle_{n}/(N-1)$ for the lowest excited states
(labeled by $n$) in the potential of fig.~\ref{fig:u_asneg}.  All
states with $n\le58$ have negative energy and $20b \leq (\langle
r_{ij}^2\rangle_{n})^{1/2} \leq 100b$, which implies that the
particles are separated more than their interaction range.  Whether
these average distances allow qualification as condensates depends on
the decay rate of these states.

When $n\ge 59$ the energies are positive and the average particle
distance now suddenly exceeds $2000 b$.  In fact we now find $\langle
r_{ij}^2\rangle \simeq 3b\sub t^2$ which approximately is obtained in
the limit of a non-interacting gas.  These states probably qualify as
condensates.  Their interaction energies are in
fig.~\ref{fig:energy_N} compared to the Gross-Pitaevskii values.  The
discontinuity at $n=58,59$ is due to the intermediate barrier.
Decreasing and eventual removal of the barrier would smear out this
abrupt change of size.  Some of the negative-energy states could then
extend very far out and in fact have sizes comparable to the trap
length.  This investigation could be repeated for the higher adiabatic
potentials, still neglecting the couplings.  The same pattern is
obtained with fewer states of small interparticle distance.

\section{Decay rates}

\label{sec:decay-rates}

\begin{figure}[htb]
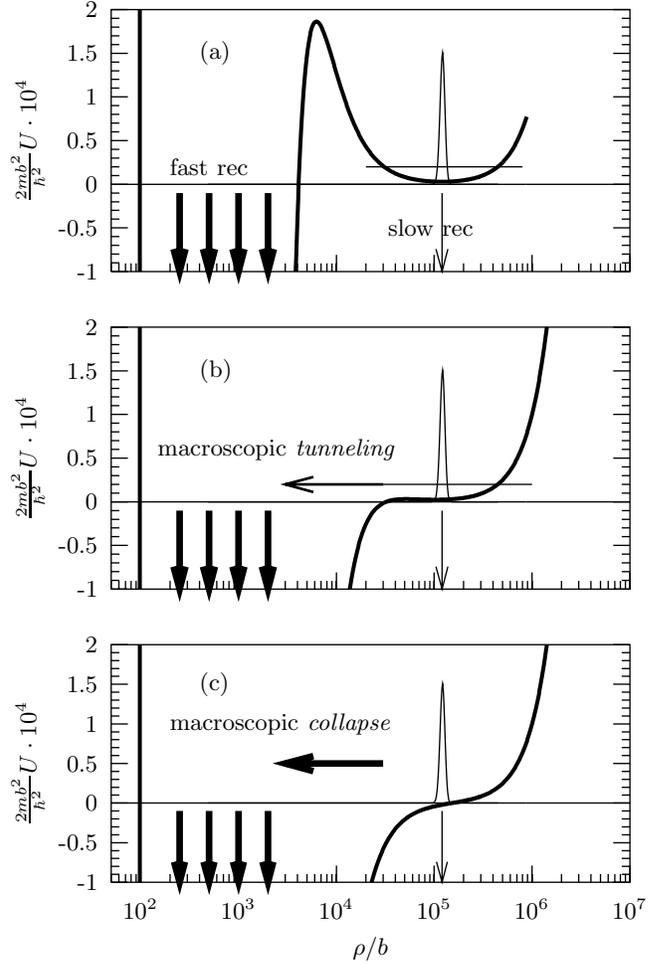

  \input{olep2fig10a.tex}
  \input{olep2fig10b.tex}
  \input{olep2fig10c.tex}
  \caption[]
  {The radial potential from the schematic model for $N=100$, $b\sub
    t/b=10^4$ and a) $a_s/b=-6$, b) $a_s/b=-50$, c) $a\sub
    s/b\to-\infty$.  The wave function is the lowest radial solution
    in the non-interacting case.  The horizontal lines in parts a) and
    b) indicate an energy level (not to scale).}
  \label{fig:cartoon}
\end{figure}

The condensate is unstable due to the neglected couplings into other
degrees of freedom.  The condensate therefore has to be located at
relatively large distances.  The decisive radial potentials are
sensitively depending on the scattering length. In
fig.~\ref{fig:cartoon} we illustrate the different behaviour by using
the angular eigenvalues parametrized through
eqs.~(\ref{eq:radial.potential}), (\ref{e19}), (\ref{e20}),
(\ref{e26}), and (\ref{e27}).  In fig.~\ref{fig:cartoon}a the
scattering length is relatively small and a large barrier separates
the outer minimum from the inner region.  By increasing the scattering
length the barrier decreases first into a relatively flat region as in
fig.~\ref{fig:cartoon}b and then disappears completely as in
fig.~\ref{fig:cartoon}c when the trap length is exceeded.  With these
potentials we can now discuss various decay processes, i.e.,
three-body recombination into dimers, macroscopic tunneling through
the barrier and macroscopic collapse after sudden removal of the
barrier.

\subsection{Three-body recombination}

Bound state dimers can be formed by a three-body process where the
third particle ensures conservation of energy and momentum.  The
number of these three-body recombination (rec) events per unit volume
and time can be estimated by the upper limit given in
\cite{nie99,bed00}:
\begin{eqnarray}
  \nu\sub{rec}=67.9\frac{\hbar |a_s|^4n^3}{m}
  \;,
\end{eqnarray}
where $n$ is the density of the gas.  This expression can be converted
into an estimate of the recombination rate for a given hyperradius
$\rho$.  With the volume $\mathcal{V} = N/n$, the relation between
density and mean distance $1/n = 4\pi \langle r_{ij}^2 \rangle ^{3/2}
/3$, and $\langle r_{ij}^2 \rangle = 2 \langle \rho ^2 \rangle /(N-1)$
obtained from eq.~(\ref{eq:rho2}), the total recombination rate
becomes
\begin{eqnarray}
 \frac{\Gamma\sub{rec}}{\hbar} = \nu\sub{rec} \mathcal{V}
 \simeq 0.5 \frac{\hbar |a_s|^4N^4}{m\bar\rho^6}
  \;,
  \label{eq:Gammarec}
\end{eqnarray}
where the mean square average $\bar\rho$ is defined as
$\bar\rho^2\equiv \langle\rho^2\rangle$.  In the spirit of the
adiabatic hyperspherical expansion method we use $\bar\rho$ as a
classical parameter.  The recombination rate increases rapidly with
decreasing $\bar\rho$, as indicated by the vertical arrows in
fig.~\ref{fig:cartoon}.

The recombination time $T\sub{rec}$ is defined by $N(t) =
N(0)\exp(-t/T\sub{rec})$, where $N(t)$ is the number of remaining
atoms.  We then obtain $\Gamma\sub{rec}/\hbar \equiv -dN/dt =
N/T\sub{rec}$ and
\begin{eqnarray}
  T\sub{rec}
  =
  \frac{N\hbar}{\Gamma\sub{rec}}
  =
  \frac{2m\bar\rho^6}{\hbar |a_s|^4 N^3}
  =
  \frac{m\bar r_{ij}^6}{4\hbar |a_s|^4}
  \;,
  \label{eq:5}
\end{eqnarray}
where we used $\bar r_{ij} = \sqrt{2/N}\bar\rho$.  The final
expression for $T\sub{rec}$ is independent of $N$.  Since the
condensate has to form in the external trap it is reasonable to define
stability against recombination by $T\sub{rec} \gg T\sub{c} \equiv
2\pi/\omega$, where $T\sub{c}$, is the oscillator time.  With
$1/\omega=mb\sub t^2/\hbar$ and eq.~(\ref{eq:5}) we get stability when
$\bar r_{ij} \gg \sqrt[6]{8\pi}|a_s|^{2/3}b\sub t^{1/3} =
d\sub{c}$, which provides $d\sub{c}$ introduced in
section~\ref{sec:defin-cond}.  In units of $b\sub t$ we obtain
\begin{eqnarray}
  \frac{d\sub{c}}{b\sub t}
  =
  \sqrt[6]{8\pi}\bigg(\frac{|a_s|}{b\sub t}\bigg)^{2/3}
  \;.
\end{eqnarray}

Thus for $|a_s|/b\sub t\lesssim1$ also $d\sub c/b\sub t\lesssim1$.
The average particle distance $\bar r_{ij}$ for a state located in the
second minimum is of the order $b\sub t$ and therefore $\bar
r_{ij}\gtrsim d\sub c$, i.e., for these states $\bar r_{ij}$ is larger
than the critical stability length $d\sub c$.  These states then
qualify as condensates.  For $^{87}$Rb atoms ($a_s \simeq 100$
a.u.) trapped in a field of $\nu \simeq 100$ Hz we obtain
$T\sub{rec}\sim7$ days.

\subsection{Macroscopic tunneling}

The second decay process is related to macroscopic tunneling through
the barrier, as indicated in fig.~\ref{fig:cartoon}b.  The present
model provides stationary eigenstates (within the allowed Hilbert
space) which by definition are time independent.  Thus, strictly the
states do not tunnel through the barrier, but an exponentially small
tail extends to small hyperradii.  All particles in this tail would
immediately recombine into molecular clusters, because the density is
very large in the inner region (both $\bar\rho$ and $r_{ij}$ are
small).  The rate of this two-step decay, i.e., tunneling through the
barrier and subsequent recombination, can be computed as the knocking
rate multiplied by the transmission coefficient, which is a measure of
the ratio of the probabilities at the turning points inside and
outside the barrier.  The rate of recombination due to macroscopic
tunneling can then be estimated semi-classically as in \cite{boh98} by
\begin{eqnarray} \label{e32}
  \frac{\Gamma\sub{tun}}{\hbar} &\simeq& \frac{N\nu} {1 + e^{2\sigma}}
  \;,\quad  \\ \label{e33}
  \nu &=& \frac{1}{2\pi}
  \sqrt{\frac{1}{m}\frac{d^2U(\rho)}{d\rho^2}\bigg|_{\rho\subsub{min}}}
  \;,\\ \label{e34}
  \sigma &=& \int_{\rho\subsub{in}}^{\rho\subsub{out}}
  d\rho\sqrt{\frac{2m}{\hbar^2} \big[U(\rho)-E\big]}
  \;,
\end{eqnarray}
where the factor $N$ is needed to give the total number of recombined
particles.  Here $\rho\sub{min}$ is the position of the second minimum
of $U$ and $\rho\sub{in}$ and $\rho\sub{out}$ are the classical
turning points of the barrier.

The barrier depends strongly on the combination $N|a_s|/b\sub t$
\cite{boh98,sor02}.  When $N|a_s|/b\sub t\ll 1$ the barrier is
large and the very small rate can be estimated through
eqs.~(\ref{e32}), (\ref{e33}), and (\ref{e34}).  The WKB action
integral is
\begin{eqnarray}
 \sigma \simeq \frac{3}{2} N \ln\bigg(\frac{b\sub t}{N|a_s|} \bigg)
  \;.
\end{eqnarray}
The barrier is absent when $N|a_s|/b\sub t \geq 0.53$. Close to,
but before reaching, this threshold of stability the WKB-exponent can
be approximated by
\begin{eqnarray}
  \sigma\simeq
  1.7 N \delta_s
  \;,\quad
  \delta_s\equiv 1-\frac{N|a_s|/b\sub t}{0.53}
  \;,
\end{eqnarray}
which is valid when $\delta_s$ is close to zero.

The barrier is observed to vanish when $N|a_s|/b\sub t \simeq
0.53$ \cite{rob01,don01}, which due to the factor of $N$ implies that
$|a_s|/b\sub t\ll 1$.  Therefore close to this threshold we have
for a condensate in the second minimum that $\bar r_{ij} \sim b\sub
t\gg d\sub c$, i.e., the three-body recombination does not limit the
stability.  In the limit $\sigma \ll 1$ we get explicitly
\begin{eqnarray}
  \frac{\Gamma\sub{rec}}{\Gamma\sub{tun}}
  \simeq
  \frac{1}{7.0 N^4}
  \ll
  1
  \label{eq:6}
\end{eqnarray}
implying that the macroscopic tunneling process dominates.  With
$\sigma\ll1$ we obtain that $\Gamma\sub{tun}/\hbar = N/T\sub{tun}
\simeq 0.5N\nu$, which for $\nu\simeq 100$ Hz corresponds to a
macroscopic tunneling time of $10$ ms.  This is much faster than the
three-body recombination time when the barrier is small
($\sigma\ll1$), i.e., $T\sub{rec} \gg T\sub{tun}$, see
eq.~(\ref{eq:6}).

\begin{figure}[htb]
  \begin{center}
    \input{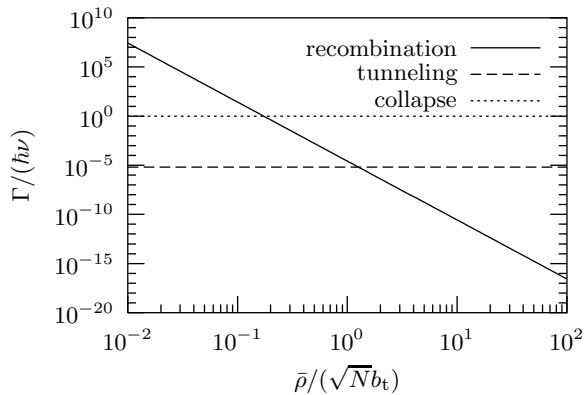}
  \end{center}
  \caption []
  {Three-body recombination rate eq.~(\ref{eq:Gammarec}) in units of
    the oscillator frequency $\nu=\omega/(2\pi)$, typically of the
    order of 10-100 Hz \cite{don01}, as a function of hyperradius for
    $N=100$, $a_s/b=-50$, $b\sub t/b=10^4$.  Shown as the
    horizontal, dashed line is the macroscopic tunneling rate
    eq.~(\ref{e32}).  Shown as the horizontal, dotted line is the
    macroscopic collapse rate eq.~(\ref{eq:collapserate}) when the
    scattering length is much larger than the trap length.}
  \label{fig:lifetimes}
\end{figure}

The three-body recombination rate is in fig.~\ref{fig:lifetimes} shown
as a function of hyperradius (solid curve) and compared with the
macroscopic tunneling rate (dashed curve) where all particles in the
condensate simultaneously disappear.  At small hyperradii the
three-body recombination rate is clearly much larger than the
macroscopic tunneling rate, whereas the opposite holds for large
hyperradii.  For the parameters in fig.~\ref{fig:lifetimes} we find
that the two time-scales are roughly equal around the second minimum
where the condensate is located.  

However, the tunneling rate depends strongly on the barrier through
the combination $N|a_s|/b\sub t$.  Varying either of the three
quantities would then move the tunneling rate up or down in
fig.~\ref{fig:lifetimes}.  For a larger barrier the condensate would
only decay by direct recombination.  For a smaller barrier macroscopic
tunneling would dominate and the condensate would decay by
``collective'' recombination of all particles in a very short time
interval.

When a few particles recombine into dimers and leave the condensate,
the remaining system is no longer in an eigenstate of the
corresponding new Hamiltonian.  An adiabatic adjustment of Hamiltonian
and wave function could then take place. Since fewer particles and
unchanged $a_s$ and $b\sub t$ means a larger barrier, the
stability against macroscopic tunneling of the new system is therefore
increased.

This stabilization by particle ``emission'' could also be the result
of the recombination in the macroscopic tunneling process if the
time-scale for recombination at the relevant small distances is longer
than the adiabatic adjustment time.  In a possible development first a
number of particles are emitted, the adjustments follow, and a larger
barrier appears which traps and stabilizes the part of the initial
wave function in the second minimum.  However, now the condensate
contains fewer particles.

\subsection{Macroscopic collapse}

These decay scenarios are open for direct experimental investigations
since the interaction can be changed in an experiment by using the
Zeeman splitting to tune to a Feshbach resonance
\cite{ino98,rob01,don01}.  An initial value of the scattering length
(corresponding to a stable condensate in the second minimum) can
almost instantaneously be changed to a value where the barrier is
removed.  The initial wave function for the condensate is now no
longer a stationary state in the new potential.

If we assume that the only excitations are the degrees of freedom
contained in the lowest new hyperspherical potential with $s$-waves, we
can use the sudden approximation and expand on the corresponding
eigenfunctions.  The most important levels are then the lowest-lying
positive energy states with energies comparable to the initial
condensate.  The time-scale for the time evolution of the initial
state in the new potential is then determined by the energy
differences between such levels.  These states of positive energy and
large spatial extension confined by the trap are roughly separated by
the oscillator quantum of energy $\hbar\omega$.  The corresponding
rate for populating smaller distances with the consequence of
immediate recombination is then crudely estimated to be
\begin{eqnarray}
  \frac{\Gamma\sub{col}}{\hbar}
  \sim
  \frac{1}{T\sub{col}}
  \sim \frac{\omega}{2\pi}
  \;.
  \label{eq:collapserate}
\end{eqnarray}
Experimentally \cite{don01} this macroscopic collapse time is verified
to be of the order $\sim 1/\omega$, typically a few milliseconds, as
given by the external trapping field.

This macroscopic collapse time is shorter than the macroscopic
tunneling time for the parameters of the system in
fig.~\ref{fig:lifetimes}.  The motion in the potential is fast or slow
compared to the recombination time for distances in the first or
second minimum, respectively. The time evolution after the sudden
removal of the barrier could then be a macroscopic collapse towards
smaller hyperradii where dimers and trimers are ``emitted'' and the
barrier begins to appear.  The part of the wave function trapped at
large distances in the second minimum can then stabilize into a
condensate with fewer particles.  The time-scale for these processes
should then be between the macroscopic collapse time and the
recombination time at the second minimum.  Possibly other time scales
due to the neglected degrees of freedom (angular momentum,
clusterization, etc.) could be present in the full study of the
dynamics of a many-boson system.

\section{Summary and conclusion}

\label{sec:conclusion}

Correlations in a system of $N$ identical bosons of low density are
described by use of a hyperspherical adiabatic expansion.  The wave
function is decomposed in additive Faddeev-Yakubovski{\u{\i}}
components, where each term is related to one pair of particles and
only $s$-waves are included.  The adiabatic potentials are only weakly
coupled and we investigate structures where only the lowest
contributes.  We use a finite-range purely attractive or purely
repulsive Gaussian interaction and extract general properties of the
lowest angular eigenvalue.  

We establish universal scaling relations for the radial potential for
arbitrary scattering length and particle number.  These scaling rules
are valid for large and intermediate distances where the particles on
average are outside the range of the interaction.  Only the
short-distance behaviour is influenced by the choice of interaction
potential.

We parametrize the model-independent part of the effective radial
potential in a simple form with an interaction part, a centrifugal
barrier term and a contribution from the external field.  This
potential diverges at small distances due to the centrifugal barrier
and at large distances due to the confining external field.  The two
minima are generally separated by a barrier.  The deepest minimum at
small to intermediate distances supports self-bound $N$-body systems
where the density is much larger than for a Bose-Einstein condensate.
The second minimum at a much larger distance allows solutions with
properties characteristic of a condensate.  We distinguish by
formulating a definition of a condensate in this context.  

We compare properties of the correlated structures with those of the
zero-range mean-field solutions.  The large-distance asymptotic
behaviour is found numerically to reproduce the mean-field result for
a zero-range interaction renormalized to give the correct scattering
length in the Born approximation.  This is remarkable since the
correct scattering length for the Gaussian potential is far from the
Born approximation.  Thus the different terms in the second-order
integro-differential equation conspire to produce this large-distance
result, which is rigorously established for three particles and on
general grounds also expected for many particles.  The choice of wave
function is then a posteriori shown to be sufficient.

The stability of the condensate is limited by decay into lower-lying
many-body cluster states reached by processes where three-body
recombination resulting in bound dimers is very prominent.  We compute
various rates of decay and discuss the time-scales involved.  The bare
three-body recombination process is strongly scattering length and
density dependent and therefore increases dramatically when the wave
packet moves from the second minimum to smaller distances.  An
intermediate barrier would only allow quantum tunneling followed by a
macroscopic collapse.  When this barrier is very small by choice of
parameters the macroscopic tunneling rate would dominate.  When the
interaction is changed during an experiment and the barrier is totally
removed the already created condensate would collapse and a number of
cluster configurations would appear.  Stability may subsequently be
automatically restored and a new condensate created with fewer
particles.

In conclusion, we have discussed properties of condensates and
extracted universal scaling relations.  We have focused on the effects
of correlations for large scattering lengths where the mean-field
approximation breaks down.  Finally we investigated time-scales for
various decay mechanisms limiting the stability of the condensate.
The parametrized potentials allow independent investigations without
the full numerical machinery.  More general $N$-body structures are
studied than the simple condensates.

\appendix

\section{Numerical details}

\label{sec:numerical-details}

The angular equation can be scaled by using the potential range $b$ as
the unit length \cite{sor02b}.  The only interaction parameter is then
the Born approximation to the scattering length in this unit $a\sub
B/b$.  The only length coordinate is then $\rho/b$.  All physical
quantities are functions of such dimensionless ratios.

The $s$-wave two-body scattering length is the node of the zero-energy
solution to the two-body Schr\"odinger-equation, i.e., $u(r) \propto
(r-a_s)$.  Table~\ref{table:as_aB} shows the scattering length $a_s$
for different potential strengths $a\sub B$, see eq.~(\ref{eq:7}).
The Born-approximation equals the correct scattering length only in
the limit of weak attraction, where the scattering length $a_s$ is
much smaller than the range of the interaction $b$.

\begin{table}[tb]
  \caption
  []
  {The scattering length $a_s$  in units of $b$  for various
    strengths of a Gaussian potential measured as $a\sub B/b$.  
    The number $\mathcal N\sub B$ indicates the number of 
    bound two-body states.}
  \begin{tabular}{|c|c|c|}
    \hline
    $a\sub B/b$ & $a_s/b$ & $\mathcal N\sub B$ \\
    \hline
    $+3.625$     & $+1.00$  & 0 \\
    $-0.35600$  & $-0.50$ & 0 \\
    $-0.50000$  & $-0.84$ & 0 \\
    $-0.551$    & $-1.00$  & 0 \\
    $-1.00000$  & $-5.98$ & 0 \\
    $-1.069$    & $-10.0$ & 0 \\
    $-1.18600$  & $-401$  & 0 \\
    $-1.18765$  & $-799$  & 0 \\
    $-1.18900$  & $-4212$ & 0 \\
    \hline
    $-1.20280$  & $+100$  & 1 \\
    $-1.33800$  & $+10.0$   & 1 \\
    $-1.500$     & $+5.29$ & 1 \\
    $-6.868$    & $-1.00$  & 1 \\
    $-7.6612$   & $-10.0$ & 1 \\
    \hline
  \end{tabular}
  \label{table:as_aB}
\end{table}

To exemplify, in experimental work $^{87}$Rb atoms with a scattering
length of $a_s=100$ a.u.~are trapped in an external trap of
frequency $\nu=100$ Hz \cite{boh98}.  Assuming an interaction range
around $b=1$ nm we obtain $a_s/b=5.29$, $b\sub t/b=1442$.  This
can be modelled by a Gaussian two-body interaction with $a\sub
B/b=-1.5$, where the lowest solution corresponds to a two-body bound
state and the next accounts for the properties of the condensate.

%\bibliographystyle{prsty}
%\bibliography{/usr/users/oles/Few-body/Skriblerier/bibliografi.bib}

\begin{thebibliography}{10}

\bibitem{bos24}
S.~N. Bose, Z. Phys. {\bf 26},  178  (1924).

\bibitem{and95}
M.~H. Anderson {\it et~al.}, Science {\bf 269},  198  (1995).

\bibitem{bra95}
C.~C. Bradley, C.~A. Sackett, J.~J. Tollett, and R.~G. Hulet, Phys. Rev. Lett.
  {\bf 75},  1687  (1995).

\bibitem{dav95}
K.~B. Davis {\it et~al.}, Phys. Rev. Lett. {\bf 75},  3969  (1995).

\bibitem{edw95}
M. Edwards and K. Burnett, Phys. Rev. A {\bf 51},  1382  (1995).

\bibitem{bay96}
G. Baym and C.~J. Pethick, Phys. Rev. Lett. {\bf 76},  6  (1996).

\bibitem{dal99}
F. Dalfovo, S. Giorgini, L.~P. Pitaevskii, and S. Stringari, Rev. Mod. Phys.
  {\bf 71},  463  (1999).

\bibitem{pet01}
C.~J. Pethick and H. Smith, {\em Bose-Einstein Condensation in Dilute Gases}
  (Cambridge University Press, Cambridge, 2001).

\bibitem{nie99}
E. Nielsen and J.~H. Macek, Phys. Rev. Lett. {\bf 83},  1566  (1999).

\bibitem{esr99}
B.~D. Esry, C.~H. Greene, and J. P.~Burke, Jr, Phys. Rev. Lett. {\bf 83},  1751
   (1999).

\bibitem{bed00}
P.~F. Bedaque, E. Braaten, and H.-W. Hammer, Phys. Rev. Lett. {\bf 85},  908
  (2000).

\bibitem{rob01}
J.~L. Roberts {\it et~al.}, Phys. Rev. Lett. {\bf 86},  4211  (2001).

\bibitem{don01}
E.~A. Donley {\it et~al.}, Nature (London) {\bf 412},  295  (2001).

\bibitem{adh02b}
S.~K. Adhikari, Phys. Rev. A {\bf 66},  013611  (2002).

\bibitem{adh02d}
S.~K. Adhikari and P. Muruganandam, J. Phys. B {\bf 35},  2831  (2002).

\bibitem{jas55}
R. Jastrow, Phys. Rev. {\bf 98},  1479  (1955).

\bibitem{cow01}
S. Cowell {\it et~al.}, Phys. Rev. Lett. {\bf 88},  210403  (2002).

\bibitem{mou01}
C.~C. Moustakidis and S.~E. Massen, Phys. Rev. A {\bf 65},  063613  (2001).

\bibitem{fad61}
L.~D. Faddeev, J. Exptl. Theoret. Phys. (U.S.S.R.) {\bf 39}, 1459 (1960) [Sov.
  Phys. JETP {\bf 12}, 1014 (1961)]  .

\bibitem{yak67}
O.~A. Yakubovski{\u{\i}}, Yad. Fiz. {\bf 5}, 1312 (1967) [Sov. J. Nucl. Phys.
  {\bf 5}, 937 (1967)]  .

\bibitem{esr99b}
B.~D. Esry and C.~H. Greene, Phys. Rev. A {\bf 60},  1451  (1999).

\bibitem{boh98}
J.~L. Bohn, B.~D. Esry, and C.~H. Greene, Phys. Rev. A {\bf 58},  584  (1998).

\bibitem{bar99a}
N. Barnea, Phys. Lett. B {\bf 446},  185  (1999).

\bibitem{blu02c}
D. Blume and C.~H. Greene, Phys. Rev. A {\bf 66},  013601  (2002).

\bibitem{jon02}
S. Jonsell, H. Heiselberg, and C.~J. Pethick, Phys. Rev. Lett. {\bf 89},
  250401  (2002).

\bibitem{sor02}
O. S{\o}rensen, D.~V. Fedorov, and A.~S. Jensen, Phys. Rev. Lett. {\bf 89},
  173002  (2002).

\bibitem{sor02b}
O. S{\o}rensen, D.~V. Fedorov, and A.~S. Jensen, Phys. Rev. A {\bf 66},  032507
   (2002).

\bibitem{sor01}
O. S{\o}rensen, D.~V. Fedorov, A.~S. Jensen, and E. Nielsen, Phys. Rev. A {\bf
  65},  051601  (2002).

\bibitem{bij40}
A. Bijl, Physica {\bf 7},  869  (1940).

\bibitem{din49}
R.~B. Dingle, Phil. Mag. {\bf 40},  573  (1949).

\bibitem{fed93}
D.~V. Fedorov and A.~S. Jensen, Phys. Rev. Lett. {\bf 71},  4103  (1993).

\bibitem{jen97}
A.~S. Jensen, E. Garrido, and D.~V. Fedorov, Few-Body Syst. {\bf 22},  193
  (1997).

\bibitem{fed01}
D.~V. Fedorov and A.~S. Jensen, Phys. Rev. A {\bf 63},  063608  (2001).

\bibitem{fed01b}
D.~V. Fedorov and A.~S. Jensen, J. Phys. A {\bf 34},  6003  (2001).

\bibitem{nie01}
E. Nielsen, D.~V. Fedorov, A.~S. Jensen, and E. Garrido, Phys. Rep. {\bf 347},
  373  (2001).

\bibitem{bul02}
A. Bulgac, Phys. Rev. Lett. {\bf 89},  050402  (2002).

\bibitem{efi70}
V. Efimov, Phys. Lett. {\bf 33B},  563  (1970).

\bibitem{pou02}
U.~V. Poulsen, Ph.D. thesis, Department of Physics and Astronomy, University of
  {\AA}rhus, Denmark, 2002.

\bibitem{ino98}
S. Inouye {\it et~al.}, Nature (London) {\bf 392},  151  (1998).

\end{thebibliography}

\end{document}